\documentclass[sigconf]{acmart}
\usepackage{booktabs}
\usepackage{tabularx}
\usepackage{graphicx}
\usepackage{hyperref}

\newcommand{\edited}[1]{\textcolor{black}{#1}}

\renewcommand\footnotetextcopyrightpermission[1]{} %

\copyrightyear{2025}
\acmYear{2025}
\setcopyright{rightsretained}
\acmConference[CSCW Companion '25]{Companion of the Computer-Supported Cooperative Work and Social Computing}{October 18--22, 2025}{Bergen, Norway}
\acmBooktitle{Companion of the Computer-Supported Cooperative Work and Social Computing (CSCW Companion '25), October 18--22, 2025, Bergen, Norway}\acmDOI{10.1145/3715070.3749261}
\acmISBN{979-8-4007-1480-1/2025/10}

\begin{document}

\title{Ethical Considerations for Observational Research in Social VR}

\author{Victoria Chang}
\affiliation{%
  \institution{University of Maryland, College of Information}
  \city{College Park}
  \state{MD}
  \country{USA}
}

\author{Caro Williams-Pierce}
\affiliation{%
  \institution{University of Maryland, College of Information}
  \city{College Park}
  \state{MD}
  \country{USA}
}
\author{Huaishu Peng}
\affiliation{%
  \institution{University of Maryland, Department of Computer Science}
  \city{College Park}
  \state{MD}
  \country{USA}
}

\author{Ge Gao}
\affiliation{%
  \institution{University of Maryland, College of Information}
  \city{College Park}
  \state{MD}
  \country{USA}
}

\begin{abstract}
Social VR introduces new ethical challenges for observational research. The current paper presents a narrative literature review of ethical considerations in observational methods, with a focus on work in HCI. We examine how unobtrusive or selectively disclosed observation is implemented in public face-to-face and social VR settings. Our review extends ethical discussions from traditional public research into the context of social VR, highlighting tensions between observer visibility, data traceability, and participant autonomy. Drawing on insights distilled from prior literature, we propose five constructive guidelines for ethical observational research in public social VR environments. Our work offers key implications for future research, addressing anticipated improvements in platform design, the management of researcher presence, and the development of community-informed consent mechanisms. 

\end{abstract}


\maketitle

\section{Background}

Observational methods have long played a critical role in sociology, anthropology, and HCI, offering qualitative insights into human behavior in public settings. When the behavior is not sensitive and does not involve vulnerable populations, such research is generally considered ethically acceptable. The American Sociological Association’s \textit{Scope of Informed Consent} states that researchers ``\textit{may conduct research in public places or use publicly available information about individuals ... without obtaining consent}'' and should ``\textit{disguise the identity of research participants ... or other recipients of their service}'' whenever confidential details arise. These guidelines have supported observational studies in parks, sidewalks, and other open-access venues.

Despite these guidelines, many researchers continue to grapple with what constitutes ethical observation—especially in covert cases where behavior is recorded without the subjects' knowledge or awareness of the researcher's presence~\cite{marzano2021covert, spicker2011covert, power1989participant}. Roulet and Marzano argue that full transparency in covert observation is often impractical, requiring researchers to make situational judgments about when and how to disclose their role. Similarly, Podschuweit~\cite{podschuweit2021covert} observes that post hoc disclosure may cause discomfort, and that ethical decisions around transparency are often negotiated in situ rather than determined by fixed rules~\cite{li2008ethics, roulet2017covert, spicker2011covert}. These reflections underscore the ethical complexity of observing public behavior, particularly when individuals are unaware they are being observed and not directly interacting with the researcher.

The aforementioned ethical ambiguities become even more pronounced in social VR. The democratization of social VR platforms has enabled researchers to conduct observational studies in virtual, immersive environments—settings that, in many ways, parallel physical public spaces. In such cases, VR users often embody avatars that carry identity cues such as usernames, movement styles, and visual customizations\edited{~\cite{vic_new}}. Many platforms also support persistent data collection~\cite{cockerton2024ethics}. These characteristics raise tensions between the seemingly ephemeral nature of virtual behavior and the lasting traceability of digital interactions. As researchers increasingly adopt fieldwork-like approaches in social VR, foundational ethical considerations around consent, disclosure, and data sensitivity must be re-examined. 

In this paper, we present a literature review that examines prior work in HCI to understand how observational methods have been implemented and what ethical norms have emerged across these domains. We then outline a set of emerging ethical challenges and propose considerations and future directions for responsible observation in social VR environments.

\section{Method}
We conducted a narrative literature review~\cite{rother2007systematic, baumeister1997writing, rumrill2001using} to examine behavioral observation in public face-to-face and social VR spaces, as well as ethical guidelines relevant to research in social VR environments. This approach enabled us to synthesize prior work and identify methodological practices and ethical considerations specific to social VR.

\subsection{Collection of papers}
To identify relevant papers, we primarily searched the ACM Digital Library for empirical studies that discussed or employed observational methods in public settings. Keyword searches were conducted within the CHI, CSCW, and UbiComp proceedings, which were selected for their established focus on VR, embodied interaction, public space research, and the ethics of observational methods. We included papers that (1) were published between 2010 and 2025, (2) employed direct behavioral observation in public spaces, and (3) discussed methodological or ethical decisions related to disclosure, consent, or data collection. Our goal was to illustrate patterns and divergences that could inform future ethical practices for observational data collection in social VR.

\textbf{Empirical studies in public face-to-face settings}.
For empirical face-to-face studies, we used the terms ``public space observation,'' ``public displays,'' ``observation,'' ``field study,'' and ``virtual reality'' to focus specifically on non-VR related papers. These searches yielded 544 papers in CHI, 86 in UbiComp, and 51 in CSCW. After reviewing abstracts and method sections, we selected a subset of \textit{nine representative papers} that employed observational methods in public face-to-face  settings for our analysis

\textbf{Empirical studies in public social VR settings.} 
For empirical social VR studies, we used the keywords ``social virtual reality'', ``observation,'' ``field study,'' and ``public.'' These searches yielded 43 papers in CHI and 6 in CSCW. We also included results from the IEEE Digital Library, which returned one relevant paper. After reviewing abstracts and method sections, we selected a subset of \textit{four representative papers} that specifically employed observational methods in public social VR settings for our analysis.

\textbf{Guidelines for the conduct of observational studies in social VR.} 
To explore existing guidelines and ethical considerations for social VR research, we first searched the ACM Digital Library using the terms ``virtual reality,'' ``observation,'' ``consent,'' and ``ethic.'' These searches returned 62 papers; however, none provided explicit guidelines for observational research in public social VR spaces. We then expanded our search to include the IEEE Digital Library, which yielded 19 papers, and APA PsycNet, which returned 5. From this broader search, we identified \textit{two papers} that specifically addressed covert observational practices within social VR environments.

\subsection{Analysis of papers}

We applied a structured set of guiding questions to analyze the collected papers:

\begin{itemize}
  \item In what type of public space did the observations occur?
  \item Was informed consent obtained, and from whom?
  \item Was the observation method described as covert, overt, ambiguous, or unreported?
  \item What type of data was collected?
\end{itemize}

For papers specifically focused on ethical guidelines, we examined how they addressed issues of informed consent, observer disclosure, and data collection practices.

\section{Findings}

The findings from our literature analysis are structured as follows: first, we synthesize empirical studies that use observational methods to collect data in both face-to-face and social VR contexts; we then zoom in on existing discussions about observational research protocols in social VR. %

\subsection{Current practices in face-to-face settings and their variations}

Observational methods have been widely used in HCI to study naturalistic behavior in public spaces, ranging from city streets to academic conferences. While methodologically similar, these studies vary in how they address ethical concerns, including informed consent, disclosure of observers, and data collection. Researchers have described covert roles using terms such as “discreet observations~\cite{chen2016led},” “covert...rapid ethnography~\cite{balestrini2016jokebox},” “in the wild field study~\cite{hespanhol2014facade, ichino2016display},” and the “shadow observer’s technique~\cite{parker2018public}.” Our analysis these variations reveal how ethics are currently interpreted in HCI observational research.

\subsubsection{Consent models and participant awareness}
One of the primary variations across these studies is how, and whether, informed consent was obtained---a central issue in the ethics of covert observational research. Some studies employed opt-in models, collecting informed consent from pre-registered or interested participants at academic conferences, although bystanders remained uninformed~\cite{mccarthy2004augmenting, chen2016led, neustaedter2016beam}.
Others relied on institutional-level consent. For example, Ichino et al. leveraged museum ticket reservation policies as consent for observing visitor behavior~\cite{ichino2016display}.
A third group of studies conducted observations in public contexts such as sidewalks and festivals without obtaining informed consent~\cite{balestrini2016jokebox, parker2018public, mueller2012glass, hespanhol2014facade} . 
They argued that the public nature of the settings permits observation, while also emphasizing steps taken to minimize identifiability in data presentation.

\subsubsection{Observer presence and disclosure}

Closely tied to informed consent is the issue of observer presence and disclosure, as this can influence participant behavior and implicate bystander privacy. Disclosure practices ranged from covert to both bystanders and any organizers, to partially disclosed in which the event organizers were aware of their activities. 
Some studies made no formal announcement of observer roles, with researchers blending into public environments~\cite{mueller2012glass, balestrini2016jokebox, parker2018public} . Others used institutional channels, such as opening plenary announcements or privacy policies tied to ticket purchases, to communicate researcher presence~\cite{neustaedter2016beam, ichino2016display}. A few studies limited observation to consenting participants but acknowledged incidental interaction with others~\cite{chen2016led, mccarthy2004augmenting}.

\subsubsection{Type of data collected during observation}

A third main area of ethical divergence considers the types of data collected. This matters in part because the sensitivity of the data, especially when collected without consent, can affect individual privacy, even in public settings. 
Approaches and data collection practices varied widely: some researchers limited their data to field notes ~\cite{mccarthy2004augmenting, hespanhol2014facade}, while others collected visual data photos, sketches, or video. In certain cases, efforts were made to obscure identities ~\cite{mueller2012glass, memarovic2015moment}; in others, researchers recorded interaction footage that included both consented participants and uninformed bystanders ~\cite{neustaedter2016beam}.

\subsubsection{Ecological validity in context}

Researchers balanced the goal of maintaining ecological validity with adherence to ethical protocols and respect for participant privacy. Practices often diverged depending on context, for example, between academic conferences (referred to as “filtered” spaces~\cite{chen2016led}) and outdoor public areas such as sidewalks. Academic conferences often involved some form of participant consent or organizer awareness, whereas public settings typically operated under the assumption that observation was permissible. Bystander consent was absent in both cases.

In summary, although observational research has been widely practiced in HCI, ethical approaches to consent, media collection, and anonymity have varied, underscoring persistent tensions between naturalistic observation and participant protection. Next, we examine how these issues manifest in social VR contexts.

\subsection{Current practices in social VR contexts and remaining ambiguities}  
Several recent studies have employed observational methods to examine social interaction in publicly accessible social VR environments. Similar to prior work in face-to-face settings, these studies involve attending live public events and documenting how participants interact in situ through avatar-mediated bodies. However, the ethical procedures across these studies are still taking shape, with researchers experimenting with their own practices as common norms continue to emerge. We examine how people implement observational research in social VR by building upon or diverging from practices established in public face-to-face settings. 

As with empirical studies in physical contexts, these ethical dimensions are especially important, as many of the studies involve some form of covert observation, described variously as “unobtrusive observations~\cite{maloney2020talking}, ” “virtual ethnography\cite{sabri2023moderating},” and “participatory observation~\cite{acena2021safe}.”

\subsubsection{Consent models and participant awareness}

The reviewed studies demonstrate differing approaches to informed consent. Maloney et al.~\cite{maloney2020talking} observed public events in AltspaceVR using a visible avatar, taking screenshots and field notes; however, their paper does not indicate whether participants or event hosts were asked for consent. Similarly, Sabri et al.~\cite{sabri2023moderating} conducted “virtual ethnography” in public VRChat spaces, and their paper does not mention any consent procedures. They participated using a visible avatar but did not recruit participants in advance or report obtaining consent from event organizers. Acena and Freeman~\cite{acena2021safe} also observed events in VRChat and Rec Room through participatory observation. While they describe engaging with participants, no informed consent is reported. They collected field data while blending into the space.

Following a slightly different approach, Chang et al. \cite{chang2024spatial} attended publicly accessible events across various social VR platforms using a visible avatar. While it was not explicitly stated whether event organizers were aware of the researcher's presence, informed consent was obtained from one participant who was already attending the event and being shadowed by the researcher. This participant was asked to review the collected material and redact any content. However, no consent was obtained from other attendees, including those who interacted with the consented participant.

Across all the studies, none reported obtaining informed consent from bystanders, and only one \cite{chang2024spatial} explicitly stated that informed consent was obtained from a specific participant at each social VR event studied. This pattern echoes the face-to-face observational studies discussed earlier, particularly those conducted outside of academic conference scenarios, where informed consent and debriefing were similarly absent.

\subsubsection{Observer presence and disclosure}
None of the reviewed studies reported disclosing the observer’s identity, purpose, or ongoing observation activities to bystanders within the social VR environment---even when directly approached. Although all researchers used visible avatar to enter events, there is no indication that they identified themselves as researchers to participants, whether verbally, through their usernames, or via avatar customization. In fact, three of the studies explicitly describe their method as designed to allow the observer to blend into the environment \cite{maloney2020talking, acena2021safe, sabri2023moderating}. 

\subsubsection{Type of data collected during observation}
The types of data collected varied across the social VR studies, ranging from field notes to recorded video. Some researchers relied on screenshots and field notes \cite{maloney2020talking, chang2024spatial, acena2021safe}, while Sabri et al. supplemented their observations with headset-based recordings \cite{sabri2023moderating}. However, most studies lacked detail about whether unaware participants were ever informed, or what anonymization steps were taken during analysis or presentation, reflecting
a range of ethical decisions regarding the sensitivity and permanence of collected data.

The above analysis highlights remaining ambiguities in ethical observational research within social VR. In particular, existing protocols for social VR research build on precedents established in face-to-face settings but introduce new challenges: avatar embodiment, discreet screenshotting and recording, and persistent data trails all heighten concerns about traceability. Although researchers often rely on the public nature of the virtual environments and events studied as justification, clear guidelines regarding observer presence and participant protection are still lacking.

\subsection{Emerging perspectives on ethics in social VR research}  
Two papers in the HCI literature explicitly engage with the ethical implications of observational research in social VR environments ~\cite{cockerton2024ethics, maloney2021virtual}. Both offer critical reflections on how publicness, embodiment, and platform design complicate long-standing assumptions about ethical observation. This discussion is especially important given that social VR platforms enable not only covert observation by researchers using avatars, but also automated---and potentially invisible---data collection of types not easily gathered in face-to-face contexts \cite{cockerton2024ethics}. 

\subsubsection{Consent models and participant awareness}
Maloney et al. \cite{maloney2021virtual} caution against equating platform-defined “publicness” with users’ or community members’ expectations. Persistent identity markers---such as usernames, voice, or avatar design---can increase the risk of identifiability. Cockerton et al. \cite{cockerton2024ethics} further argue that platform design complicates the consent process, especially when users lack visibility into or control over what is being recorded and how. Although these researchers acknowledge the difficulty of implementing opt-out mechanisms, they stress that the absence of such options may violate the principle of respect for participants.

\subsubsection{Observer presence and disclosure}
Maloney et al. \cite{maloney2021virtual} highlight the importance of researcher presence and disclosure in social VR during data collection. Even when researchers are visually present through avatars, they may blend in as regular attendees at virtual events, creating uncertainty about who is observing and for what purpose. At the same time, disclosure can itself alter behavior, potentially undermining the naturalism of the observation. Cockerton et al. \cite{cockerton2024ethics} offers a similar perspective, adding that spatial and social positioning within virtual environments can carry implicit meanings. A researcher’s proximity to others, gaze behavior, or avatar appearance may all influence how they are perceived. Yet these cues are often ambiguous, especially in platforms that lack explicit tools for communicating observer intent. Both papers recommend that researchers avoid fabricating identities during data collection and emphasize the importance of using anonymized identifiers in both data collection and presentation, particularly when working with vulnerable populations.

\subsubsection{Type and sensitivity of data collected}

Both papers emphasize that data collected in social VR---such as position logs, screenshots, voice recordings, gestures, and even more discrete biometric or neural signals---can be more identifying than comparable data in physical public settings. Maloney et al. warn that these digital traces heighten the risk of re-identification. Cockerton et al. point out that automated logging and other default platform-level data collection mechanisms can threaten transparency and undermine user control. These concerns underscore new ethical demands that are not fully addressed by traditional research practices and guidelines.

\section{Discussion and Proposed Guidelines}

Our  analysis of the literature reveals that ethical guidelines for VR-based observational research are still evolving. While observation as a research method has long been used in public face-to-face settings, its application to social VR introduces additional challenges related to the choice of consent approaches, the design of  disclosure protocols, and the management of data traceability. Future research should continue to investigate how to balance ecological validity, transparency, and participant rights in digital public spaces. Below, we proposed several key guidelines to support and enrich this ongoing discussion.

\textbf{Guideline 1: Develop adaptive consent models for observational research in social VR.}  
Building on suggestions by Cockerton et al.~\cite{cockerton2024ethics}, future research should explore dynamic, platform-integrated consent mechanisms. These could include opt-in settings or event-triggered disclosures that notify users of ongoing research in the VR environments they are participating in. Visual indicators (e.g., avatar icons) or automatic exclusion from media capture could be implemented to better align data collection practices with individual preferences and platform policies. In structured VR settings, such as virtual conferences, researchers could issue pre-event notices to inform potential attendees of the activities in which observational research is taking place. This approach can help various stakeholder groups establish joint expectations while preserving research integrity.

\textbf{Guideline 2: Address bystander data and enable opt-out options in social VR environments.}  
Public-space research has not not traditionally require individual consent with everyone. However,  future studies in social VR should explore mechanisms that allow bystanders to opt out. This is especially important given that data collected in digital environments, such as gaze and body movement tracking, can go beyond what is observable in physical settings and, as a result, may not be transparent to bystanders or even to consented participants.

\textbf{Guideline 3: Consider the ethics of persistent and reusable data collected from social VR users.}  
Unlike physical spaces, social VR allows human behaviors to be recorded, stored, and replayed indefinitely. This raises concerns about re-identifiability, long-term data use, and consent that extends beyond the lifecycle of a given research project. Future research should develop robust protocols for data retention and secondary analysis to ensure that social VR users can retain control over their digital traces.

\textbf{Guideline 4: Collaborate with social VR users and stakeholders to develop community-informed ethical framework.}  
Ethical research in social VR should be guided by norms and expectations of the communities being studied. Researchers must collaborate with platform developers and user groups to co-create ethical frameworks that reflect both scientific rigor and social responsibility. Clear and accessible scaffolding should be provided to help individuals understand each platform's terms of service and built-in data practices, ensuring their  expectations and rights are effective communicated.

\textbf{Guideline 5: Balance ecological validity with appropriate researcher disclosure in social VR.}  
Full disclosure of the researcher's identity in observational studies can sometimes disrupt the naturalistic behavior of participants. Future studies should explore context-sensitive disclosure strategies. For example, researchers should better be clearly identifiable in smaller, community-driven events, while maintaining a lighter presence in larger, unstructured spaces. As Chen and Abouzied note, virtual conferences and similar settings are often filtered spaces with different privacy expectations than transient public areas (e.g., city squares) ~\cite{chen2016led}. This implies that researchers' decisions about when and how to disclose their identity in social VR should be calibrated to the perceived publicness of the virtual space by those present in it.

\nocite{*}
\bibliographystyle{ACM-Reference-Format}
\bibliography{references}

\end{document}